\documentstyle[12pt,aaspp4,tighten]{article}
\begin{document}

\newcommand{\msun}{M{$_{\odot}$}} 
\newcommand{\lsun}{L{$_{\odot}$}}
\newcommand{\rsun}{R{$_{\odot}$}}
\newcommand{\nduehp}{N$_2$H$^+$(1--0)}
\newcommand{\methduall}{CH$_3$OH(2$_{-1}$--1$_{-1}$)E2~\& (2$_0$--1$_0$)A$^+$}
\newcommand{\methcqall}{CH$_3$OH(5$_{-1}$--4$_{-1}$)E2~\& (5$_0$--4$_0$)A$^+$}
\newcommand{\methdu}{CH$_3$OH(2$_{k}$--1$_{k}$)}
\newcommand{\methcq}{CH$_3$OH(5$_{k}$--4$_{k}$)}
\newcommand{\lsim}{\;\lower.6ex\hbox{$\sim$}\kern-7.75pt\raise.65ex\hbox{$<$}\;}
\newcommand{\gsim}{\;\lower.6ex\hbox{$\sim$}\kern-7.75pt\raise.65ex\hbox{$>$}\;}

\title{Star formation in clusters: early sub-clustering in the Serpens core}

\author{Leonardo Testi\altaffilmark{1,2,5}, Anneila I. Sargent\altaffilmark{2},
        Luca Olmi\altaffilmark{3} and Joseph S. Onello\altaffilmark{4}}

\altaffiltext{1}{Osservatorio Astrofisico di Arcetri, Largo E.~Fermi 5,
I-50125 Firenze, Italy}
\altaffiltext{2}{Division of Physics, Mathematics and Astronomy, California 
Institute of Technology, MS~105-24, Pasadena, CA~91125, USA}
\altaffiltext{3}{LMT Project and FCRAO, University of Massachusetts,
630 L.G.R.C., Amherst, MA 01003, USA}
\altaffiltext{4}{Department of Physics, State University of New York, 
Cortland, NY 13045, USA}
\altaffiltext{5}{ltesti@arcetri.astro.it}

\begin{abstract}

We present high resolution interferometric and single dish observations of
molecular gas in the
Serpens cluster-forming core. 
Star formation does not appear to be homogeneous throughout the core,
but is localised in spatially- and kinematically-separated
sub-clusters. The stellar (or proto-stellar) density in each of the
sub-clusters is much higher than the mean for the entire Serpens
cluster. This is the first observational evidence for the hierarchical
fragmentation of proto-cluster cores suggested by cluster formation
models.

\end{abstract}


\section{Introduction}
\label{sintro}

It is generally accepted that most stars are born in clusters (cf. 
Clarke et al.~\cite{CBH00}). The way in which clusters form 
and evolve is therefore likely to influence the distribution of masses 
for stars in the field, the initial mass function, IMF (Salpeter~\cite{S55}; 
Scalo~\cite{S86}). In very young, embedded clusters the distribution of stellar
masses is often similar to the IMF (Palla \& Stahler~\cite{PS99};
Meyer et al.~\cite{Mea00}), and the mass spectra of  prestellar
and protostellar condensations in the Serpens and $\rho$--Ophiuchi 
cluster-forming cores are also consistent with the IMF (Testi \&
Sargent~\cite{TS98}, hereafter TS98; Motte et al.~\cite{MAN98}).

It has been suggested that within stellar clusters star formation 
occurs preferentially in sub-clusters where the stellar density is 
much enhanced (Clarke et al.~\cite{CBH00}).
This has important implications
for cluster evolution. For example, more massive stars could be produced 
by coalescence (Stahler et al.~\cite{SPH00}). To date, there is 
little evidence for sub-clustering in the Orion Nebula Cluster 
(Bate et al.~\cite{BCM98}) or in the smaller clusters around intermediate mass
pre-main sequence 
stars (Testi et al.~\cite{TPN99}).
However, models advocate sub-clusters with a much higher stellar density at the
time of formation
(Bonnell et al.~\cite{Bea98}). It is therefore important to establish
if sub-clustering is present in the very youngest clusters. 
The mean stellar densities and the stellar to gas mass ratio in such
sub-clusters can provide critical observational constraints on
coalescence, competitive accretion and binary evolution models (Bonnell
et al.~\cite{Bea97}; \cite{Bea98}; Kroupa~\cite{Kea99}).

The Serpens molecular cloud, at $\sim310$~pc (de~Lara et al.~\cite{dLea91}),
is one of the most active nearby cluster-forming cores. Inside the
500--1500~M$_\odot$, $\sim$0.6~pc diameter
cloud of molecular gas (White et al.~\cite{WCE95})
is a young protocluster comprising about one hundred embedded 
young stellar objects (YSOs), protostars and prestellar clumps
(Strom et al.~\cite{SVS76}; Eiroa \& Casali~\cite{EC92}; Giovannetti
et al.~\cite{Gea98}; Kaas~\cite{K99}; Casali et al.~\cite{CED93};
Hurt \& Barsony~\cite{HB96}; TS98).
Numerous jets and molecular outflows 
have also been detected (Rodr\'{\i}guez et al.~\cite{Rea89}; White et 
al~\cite{WCE95}; Eiroa et al.~\cite{Eea97}; Herbst et al.~\cite{HBR97};
Wolf-Chase et al.~\cite{WCea98}; Davis et al.~\cite{Dea99}; Hodapp~\cite{H99};
Hogerheijde et al.~\cite{H99}).
The total estimated mass of the YSOs, protostars, and prestellar clumps
is in the range 
40--80~M$_\odot$ (Giovannetti et al.~\cite{Gea98}; TS98), implying
an overall star formation efficiency of 2--5\%, similar to
most nearby molecular clouds.
The proto-cluster radius, $\sim 0.2$~pc,
and mean stellar density, $\sim 400$-$800$~stars/pc$^3$, are typical of very
young embedded clusters (Testi et al.~\cite{TPN99}), making this 
an ideal laboratory for studying early cluster formation processes.
Here, we present wide field, high resolution, aperture synthesis and single
dish millimeter-wave molecular line observations of the Serpens core which
support the concept of at-birth sub-clustering.


\section{Observations and results}
\label{obser}

Owens Valley Radio Observatory (OVRO) millimeter wave array mosaic
observations of the $5.5^\prime\times 5.5^\prime$ inner region of
the Serpens molecular core in the CS(2--1) transition at 97.98~GHz
were obtained at the same time as the 
3~mm continuum data described by TS98. Details of the observations, data
reduction, and mosaic construction are described elsewhere (TS98; Testi 
\& Sargent~\cite{TS00}). Spectral resolution was $0.4$~km$/$s over
a 24~km$/$s bandwidth centered at $V_{\rm LSR}=$8~km$/$s. The synthesized
beam is $5^{\prime\prime}\!.5\times 4^{\prime\prime}\!.3$ (FWHM),
and the noise level in each channel of the final cleaned mosaic cube is
$\sim 140$~mJy$/$beam.
During spring/fall~1998, additional observations of SMM4, S68N and the CS6, CS7
and CS8 features identified in Figure~\ref{fcsmap}a were acquired (Testi \& 
Sargent~\cite{TS00}). The digital correlator was configured so that
the N$_2$H$^+$(1--0) transition at 93.2~GHz and the 
CH$_3$OH(2$_{-1}$--1$_{-1}$)E2 and (2$_0$--1$_0$)A$^+$ transitions
at 96.7~GHz were detected simultaneously in the
lower and upper sidebands respectively. Spectral resolution was 0.4~km/s
over a 25~km/s band.
Calibration was carried out as in TS98.
The datasets were imaged using the AIPS IMAGR task.

A $10'\times10'$ region of the  Serpens cloud core, centered on
$\alpha=18^{\rm h}27^{\rm m}20^{\rm s}$, $\delta=1^{\circ}12'30''$
(B1950.0), was also mapped in the N$_2$H$^+$(1--0) line in March and
October 1999, using the 13.7-m telescope of the Five College Radio
Astronomy Observatory\footnote{The Five College Radio Astronomy
Observatory is operated with support from the National Science
Foundation and with permission of the Metropolitan District Commission}
(FCRAO) 
and the SEQUOIA focal plane array. For the one beam sampled maps, the half
power beam width (HPBW) was $\sim$52$^{\prime\prime}$.  Low noise InP
MMIC-based amplifiers resulted in mean receiver temperature of 70~K
(SSB), and system temperatures of 160--250~K.  The spectrometer was an
autocorrelator with 24~kHz (0.077~km/s) spectral resolution and 20~MHz
bandwidth.  Typical integration times were 5 to 15 minutes in frequency
switching mode, with a throw of 8~MHz. The main beam efficiency, 
to convert the antenna temperature to 
brightness temperature, $\eta_{\rm mb}$, is $0.51$, and the final rms
was $\sim$0.05~K (T$_{mb}$).

\subsection{OVRO maps}

In Figures~\ref{fcsmap}a and \ref{fcsmap}b we show the CS(2--1)
integrated intensity mosaic
overlaid on our 3mm continuum map (TS98) and the CS(2--1) 
mean velocity mosaic, respectively. In Fig.~\ref{fcsmap}a some of the 
brightest CS(2--1) features are labelled to simplify discussion;
the location of the NIR cluster 
surrounding SVS-2 and SVS-20 (Strom et al.~\cite{SVS76}) is also indicated.
In Fig.~\ref{fcsmap}b, the orientations of the
jets emanating from SMM1, SMM3, SMM4, and A3, as well as the NIR reflection
nebula associated with SMM5 are marked.

At most, 20\%\ of the flux detected in single dish maps of the
optically thick and extended CS(2--1) emission (cf. McMullin et
al.~\cite{MMea94}) is recovered. There is therefore little point in
comparing the CS and the optically thin 3~mm emission (TS98). However, the
molecular line observations provide information about the outflows
emanating from the various millimeter sources; the abundance of CS can
increase by a factor of almost 100 in outflows (Bachiller \& P\'erez
Guti\'errez~\cite{BPG97}), and most of the clumps in
Figure~\ref{fcsmap}a are likely to be compact enhancements in shocked
regions. In most cases the CS linewidths are relatively broad,
$\ge$2--3~km/s, and mean velocities differ by 3-4~km/s from the
systemic velocity of the Serpens core (Figure~\ref{fcsmap}b).  Strong
and broad CH$_3$OH emission, typical of shocked material, was also seen
in our followup observations (Figure~2), and coincides with the CS
features. By contrast N$_2$H$^+$ emission is spatially coincident with
the continuum sources, as expected for a tracer of cores and
envelopes.

\subsection{FCRAO N$_2$H$^+$ map}

Figure~3 shows the FCRAO/SEQUOIA large scale channel maps of the
N$_2$H$^+$(1$_{01}$--0$_{12}$) isolated hyperfine component.  Four
cores, labelled A, B, C, and D in order of increasing V$_{LSR}$, and an
extended, lower surface brightness ``spur'' can be identified. The spur
stretches from the NW (C, D) to SE (A, B) cores and continues beyond
these.  Cores A and B are spatially associated with the SMM3/SMM4
region, while C and D encompass the SMM1/S68N region.  Spectra at the
emission peaks in A, B, C, and D are also presented in Fig.~3. These
positions are 
given in Table~1 as offsets from $\alpha=18^{\rm h}27^{\rm m}20^{\rm s}$, 
$\delta=1^{\circ}12'30''$. Also listed are the V$_{LSR}$, line width,
$\Delta$V, and virial mass, M$_{vir}$, for each core. 
The fraction of the total N$_2$H$^+$ flux observed with the array
is f$_{OVRO}$; M$_d$ is the total mass of 3~mm continuum cores within each 
gaseous core (TS98), and M$_\star$ is the total mass of YSOs, derived assuming
a mean stellar mass of $0.3$~M$_\odot$ (Giovannetti et al.~\cite{Gea98};
Kaas~\cite{K99}).
The FCRAO and OVRO values of $\Delta$V and V$_{LSR}$ are
consistent, suggesting that the array is detecting
a fraction of the emission from extended envelopes, rather than the 
compact cores found by TS98.

\subsection{Identification of outflow sources}

The powering sources of the CS outflows were identified from published
optical, infrared and millimeter observations, and are represented by
open diamonds in Figure~1.  Features CS3 and CS4 as well as CH$_3$OH
emission (Fig.~2) 
suggest a compact outflow at
p.a.$\sim$140$^\circ$, centered on the S68N continuum source (cf.
Wolf-Chase et al.~\cite{WCea98}). CS2 could either be an extension
of the S68N flow or the counterflow of the H$_2$ jet
from mm source A3 (Hodapp~\cite{H99}; TS98).  
Based on its alignment with an H$_2$ jet
(Hodapp~\cite{H99}) and the locations of HH460 and a CO(2--1) bullet
(Davis et al.~\cite{Dea99}), we associate CS1 with an
outflow from SMM1. CS6, CS7, and CS8 probably mark the
interaction of the counterflow from SMM1 with the B molecular core. All
these outflows are oriented along p.a.$\sim$140$^\circ$. A
double-peaked CS component reported by Williams \& Myers~(\cite{WM99})
at the position of CS1 was not detected; it is probably extended and
resolved out in our observations.

In the south-east, CS9 and CS10 are oriented like CH$_3$OH (Fig.~2) and
H$_2$ knots near SMM4 (Eiroa et al.~\cite{Eea97}) along
p.a.$\sim$180$^\circ$, and are probably part of an outflow centered on
that source (c.f. Hogerheijde et al.~\cite{Hea99}).
Likewise CS12 is elongated along p.a.$\sim$170$^\circ$ as
are a chain of H$_2$ knots associated with SMM3 (Herbst et
al.~\cite{HBR97}). Davis et al.~(\cite{Dea99}) noted that two Herbig-Haro
objects and SMM3 are aligned along 
a position angle which is almost orthogonal to the chain of H$_2$
knots.  However, the kinematic properties of the knots suggest no
physical connection with the HH objects.


\section{Discussion}

Our new interferometer and single-dish maps of the Serpens core
indicate sub-clustering at an early epoch of cluster formation.  Three
separate properties argue for sub-clustering, spatial segregation,
outflow orientations, and circumcluster gas kinematics.
Approximately one third of the near infrared cluster members are
concentrated in a $\sim$0.1~pc radius region surrounding SVS-2 and
SVS-20 (Giovannetti et al.~\cite{Gea98}; Kaas~\cite{K99}), while the
millimeter and sub-millimeter sources are largely concentrated in the
SE (A, B) and NW (C, D) fragments (TS98; Davis et al.~\cite{Dea99}).
The orientations of the outflows observed in the SE and NW are
quite different. In the NW, all three flows are oriented along
p.a.$\sim$140$^\circ$, as is the near infrared reflection nebula
centered on SMM5 (Kaas~\cite{K99}; see also Figure~\ref{fcsmap}b).  In
the SE, the two outflows are aligned approximately north-south, with
mean p.a.$\sim$175$^\circ$. In Figure~3, the NW and
SE sub-clusters are embedded in discrete \nduehp\ clumps, separated in
velocity by $\sim$1~km/s (Table~1). Each clump
comprises two cores: peak velocities of cores A and B differ by only
0.5~km/s ($\sim 1/2$~$\Delta$V), C and D by 0.1~km/s ($\sim 1/10$~$\Delta$V).
Thus the spatially distinct clumps are also kinematically separated, while
there is reasonable internal velocity coherence.

It appears that the Serpens core encompasses at least three
sub-structures -- the NW and SE sub-clusters and the NIR cluster. Star
formation is currently occurring simultaneously in the NW and SE
sub-clusters, both of which contain roughly equal fractions of
prestellar, protostellar and infrared sources. The NIR cluster is
probably more evolved but we see no evidence of the progressive pattern
of star formation proposed by Casali et al.~(\cite{CED93}).  The
kinematics and outflow orientations indicate that each subcluster
originated in a separate fragment of the cloud which subsequently
fragmented into the smaller cores seen in the \nduehp\ maps. Within
these are the 3~mm continuum cores that are likely progenitors of
single stellar systems (TS98). Taken together these observations are consistent
with either the hierarchical fragmentation picture advocated by
Elmegreen~(\cite{E97}; \cite{E99}) or the spontaneous fragmentation
suggested by Myers~(\cite{M98}) for the formation of stellar clusters,
and provide the first observational support for hierarchical
fragmentation within a cluster-forming core. We note that while the
mean stellar/protostellar density of the entire Serpens core is
$\sim$400-800~stars/pc$^3$, most of the proto-cluster members are
within the three subclusters where densities reach 2000-4000~stars/pc$^3$.
It is very likely that, within a few million years, the cluster will
evolve to a size and mean density very similar to those of embedded
clusters around intermediate mass stars.

\smallskip
\noindent
{\bf Acknowledgements:}  
We thank Cathie Clarke and the referee, Paul Ho, for comments which
much improved this paper.
The Owens Valley
millimeter-wave array is supported by NSF grant AST-96-13717.
Research on young star and disk systems is also supported by the
{\it Norris Planetary Origins Project} and NASA's {\it Origins of
Solar Systems} program (through grant NAGW--4030). 
The FCRAO observations were supported by NSF grant AST-97-25951.
JSO thanks the Cornell University Department of Astronomy
for continuing support and warm hospitality.

%
%
%

%
%
\begin{deluxetable}{lcccccccc}
\footnotesize
\tablecaption{N$_2$H$^+$(1--0) Cores Parameters\label{tcores}}
\tablewidth{0pt}\tablehead{\colhead{Name}&
     \colhead{{$\Delta\alpha$}}&\colhead{{$\Delta\delta$}}&
     \colhead{V$_{LSR}$}&\colhead{$\Delta$V}&
     \colhead{M$_{vir}$}&\colhead{f$_{OVRO}$}&
     \colhead{M$_d$}&\colhead{M$_\star$}\\
     \colhead{}&\colhead{(1950)}&\colhead{(1950)}&
     \colhead{(km/s)}&\colhead{(km/s)}&
     \colhead{(M$_\odot$)}&\colhead{}&
     \colhead{(M$_\odot$)}&\colhead{(M$_\odot$)}}
\startdata
A    & 133 & -177 & 7.15 & 0.7 & -- & -- & -- & --\\
B    & 88  & -44  & 7.68 & 1.2 & 30 & $\sim$20\% & 12 & 7\\
C    & -44 & 44   & 8.35 & 1.1 & 18 & -- & 14 & 2\\
D    & -44 & 133  & 8.49 & 0.8 & 9 & $\le$5\% & 5.5 & 2\\
Spur & -- & -- & 8.32 & 0.7 & -- & -- & -- & --\\
\enddata
\end{deluxetable}

\clearpage

%
%
\plotfiddle{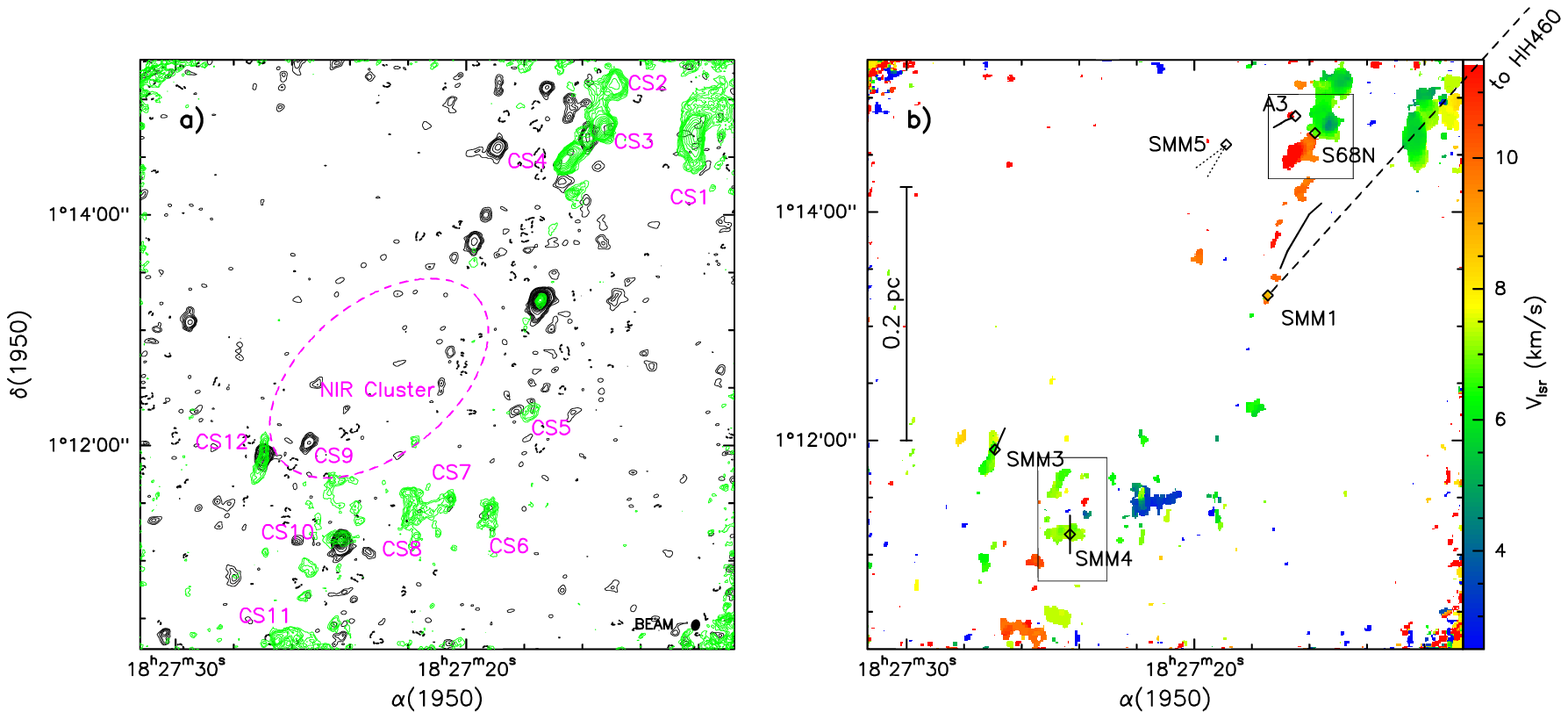}{7.5cm}{0}{105}{105}{-340}{-260}
\figcaption[fcsmos.ps]{\label{fcsmap} a) Integrated intensity contours for
CS(2--1) emission in the Serpens core are shown in green, overlaid on the 
3~mm continuum map (TS98). Contour levels begin at 0.9~Jy\,km/s/beam
(3$\sigma$) and are spaced by 0.3~Jy\,km/s/beam to 3~Jy\,km/s/beam
and thereafter by 1~Jy\,km/s/beam. The dashed ellipse represents the
NIR cluster (Kaas~\cite{K99}). The CS features discussed in the text are
marked CS1 to CS12.
b) Color-coded map of the mean velocity variations across the CS structures 
of Fig.~\ref{fcsmap}a. Known outflow sources, represented by open
diamonds, are labelled and solid lines indicate the orientations of 
observed jets or, for SMM5, reflection nebulosity. The dashed line from SMM1 
indicates the direction to HH460.
The two rectangles mark the areas shown in Figure~2. 
Increased noise in the upper left and lower right corners of both images
is due to loss of sensitivity at the edge of the mosaic.
}
\clearpage

\plotfiddle{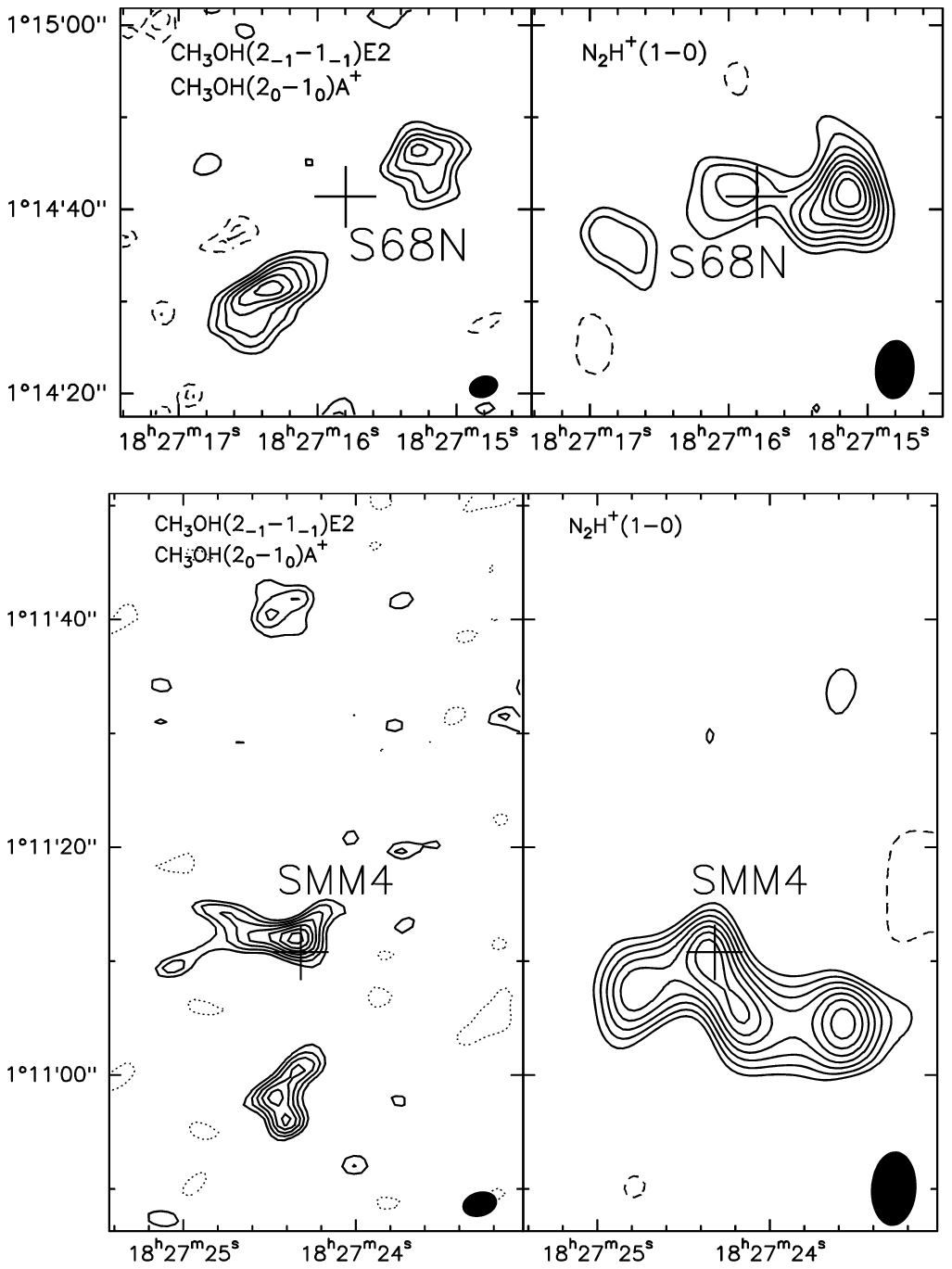}{8.5cm}{0}{60}{60}{-90}{-94}
\figcaption[flines.ps]{\label{flines} 
Integrated intensity maps of \methdu\ and
\nduehp\ for S68N and SMM4 the two regions indicated in Figure~\ref{fcsmap}b.
Synthesised beams
(FWHM) are shown by black ellipses in the lower right corners. Contour levels 
start at 3$\sigma$ and are spaced by 1$\sigma$ (0.3~Jy\,km/s/beam
for S68N and 0.2~Jy\,km/s/beam for SMM4).
}
\clearpage

\plotfiddle{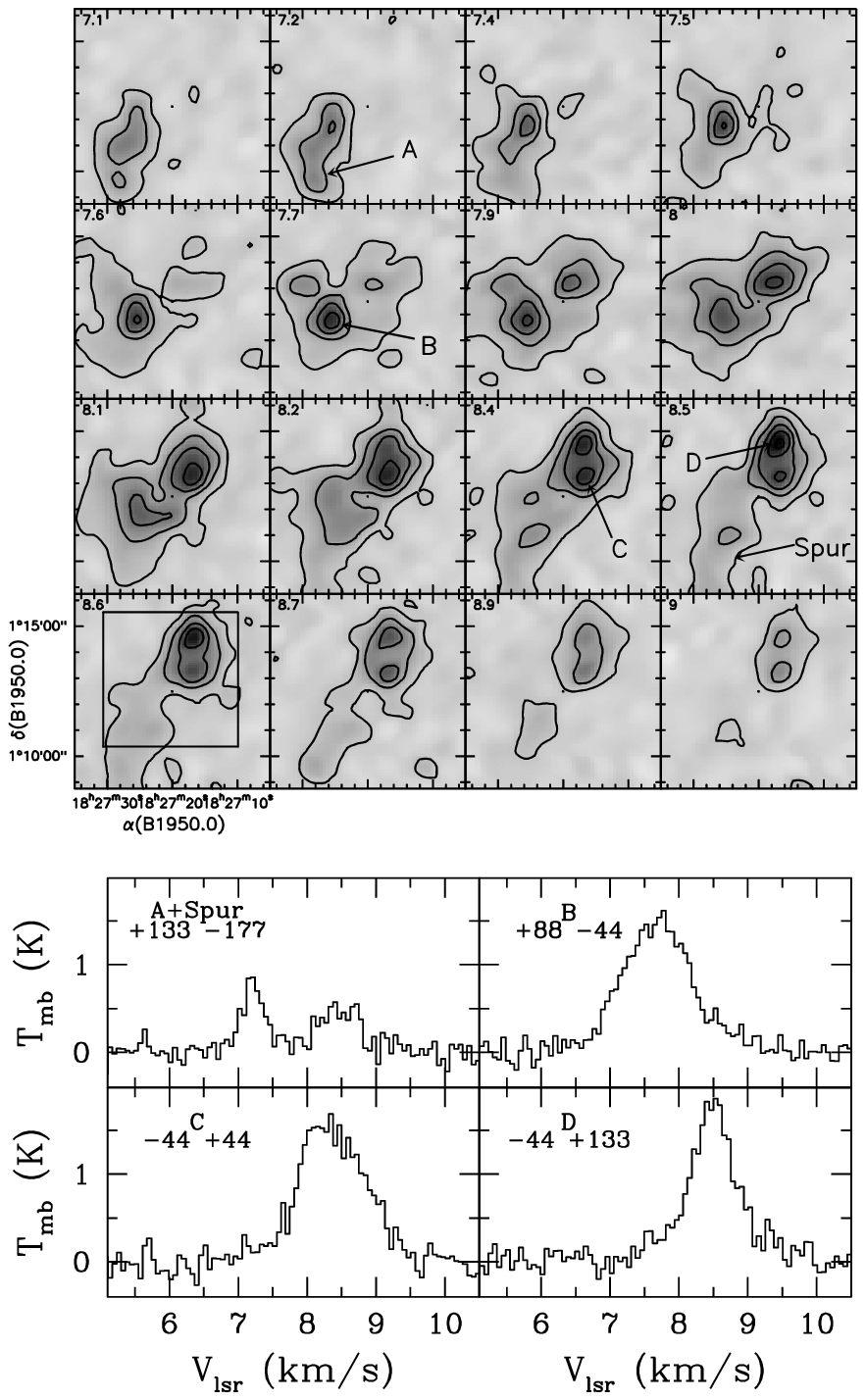}{11.5cm}{0}{90}{90}{-300}{-110}
\figcaption[fig3.ps]{\label{ffcrao} 
Top panel:
FCRAO/SEQUOIA N$_2$H$^+$(1$_{01}$--0$_{12}$)
channel maps. The five main kinematical structures, A, B, C, D,
and the ``Spur'' are each labelled at their approximate V$_{LSR}$,
which is given in the top-left corner of each panel.
Contour levels are: 0.18 to 2.18 by 0.36~K.
The square in the bottom left channel shows the
area mapped at OVRO.
Bottom panel: N$_2$H$^+$(1$_{01}$--0$_{12}$) spectra
observed towards the four cores A, B, C, and D, as defined in Fig.~3. 
In each panel the offset 
in arcsec from position $\alpha=18^{\rm h}27^{\rm m}20^{\rm s}$,
$\delta=1^{\circ}12'30''$ (B1950.0) is given.
}

\end{document}